\documentclass[a4paper,journal]{IEEEtran}
\usepackage{amsfonts}
\usepackage[fleqn]{amsmath}
\usepackage{algorithmic}
\usepackage{array}
\usepackage[caption=false,font=normalsize,labelfont=sf,textfont=sf]{subfig}
\usepackage{textcomp}
\usepackage{stfloats}
\usepackage{url}
\usepackage{verbatim}
\usepackage{graphicx}
\def\BibTeX{{\rm B\kern-.05em{\sc i\kern-.025em b}\kern-.08em
    T\kern-.1667em\lower.7ex\hbox{E}\kern-.125emX}}
\usepackage{balance}

\usepackage{cite}
\usepackage{multirow}
\usepackage{tabularx,booktabs}

\usepackage{ragged2e}

\usepackage[flushleft]{threeparttable}

\usepackage{siunitx}
\usepackage{enumerate}

\usepackage{mathtools}
\usepackage{url}

\usepackage{color,soul}

\usepackage{xcolor}

\begin{document}
\title{Extension of Simple and Accurate Inductance Estimation for Rectangular Planar Windings}
\author{\IEEEauthorblockN{Theofilos Papadopoulos, \textit{Student Member, IEEE},
		Antonios Antonopoulos, \textit{Member, IEEE}}
%		\IEEEauthorblockA{\\School of ECE,
%		NTUA, Greece\\
%		Email: teopap@mail.ntua.gr}
%		\thanks{quick footnote for comments}
	}

%\markboth{Journal of \LaTeX\ Class Files,~Vol.~18, No.~9, September~2020}%
%{How to Use the IEEEtran \LaTeX \ Templates}

\maketitle

\begin{abstract}
This paper proposes a method to generalize the equations estimating the inductance of square-shape planar windings to rectangle shape. This is done by utilizing the optimal $p$-norm of the Generalized Mean Value or Power Mean (PM). Three well-established equations with verified accuracy are examined, namely Wheeler’s, Rosa’s, and the Monomial, which by definition consider only regular polygons. One critical parameter of the original equations is the outer-side length of the winding, which for the rectangle case, can be substituted by the PM of the two outer-side lengths, without the need for any further modifications. A methodology to select the optimal $p$-norm for the PM is presented in terms of achieving the best accuracy for this estimation. The selection of the optimal $p$ is based on results from datasets containing more than 2600 simulations of different rectangle-shaped windings. Finally, the estimation accuracy is verified by laboratory measurements for a selection of planar inductors. 
\end{abstract}


\section{Introduction}
%\IEEEPARstart{W}{elcome} to the updated and simplified documentation to using the IEEEtran \LaTeX \ class file. The IEEE has examined hundreds of author submissions using this 

Developments in solid-state technology have enabled the construction of high-power and high-frequency converters. These properties allow for significant volume reduction, combined with very high efficiency figures. Achieving high-power-density designs, however, is not just a matter of squeezing more power out of switching devices, but also of the appropriate spatial planning of both the active and passive components of the system. In this sense, all surrounding components must be adapted to enable optimal converter performance. For the design of magnetic components, planar windings (PWs) offer an attractive solution, especially in designs where the converter volume is critical. Low power PWs were first developed for RF applications; however, nowadays PWs are emerging in the high-power area as well, offering several advantages compared to conventional windings for industrial applications.

PWs are the subject of extensive studies in the literature and a brief review of several aspects related to them can be found in \cite{andersen} and \cite{andersen2}. Their utilization in inductors or transformers, with or without the existence of a ferrite core\cite{tria16}, constitutes a competitive alternative to conventional concentric, solenoid windings, maintaining the advantage of high reliability. PWs extend mainly in one plane, providing low-profile magnetic components with high power density capability. Furthermore, the simplicity of their manufacturing process 
and the ability of mass production, without the need of winding-wrapping machinery, provides windings with well-predetermined inductance values \cite{fletcher}, and reduces the manufacturing cost. Compared to conventional designs, PWs provide good thermal characteristics and can operate at lower temperatures for the same amount of power, due to their larger surface area  \cite{thermal1}.
Furthermore, they are less susceptible to eddy current and proximity losses, due to copper's low height (approx. 35 \textmu m for most applications), but they can also be designed as planar-Litz windings for frequencies above which eddy currents start to play a significant role \cite{Litz21}.

Several applications can take advantage of the aforementioned properties, such as data centers, wireless/inductive power transfer systems\cite{liu21,IPT,emin} and electrical vehicles \cite{wpt1, ev1}, where low-cost and low-profile magnetic components are extremely important considering the spatial limitations. Niche applications have also been proposed in literature \cite{spro}, like coreless long-distance isolated power transfer for auxiliary supplies in medium-voltage converters. In all the aforementioned applications, precise knowledge of the magnetizing and leakage inductances of the magnetic component is of paramount importance \cite{margueron1, buttay}. Linking the physical dimensions to the magnetic properties is desirable for the converter's designer, in order to combine spatial planing with electrical and magnetic design\cite{thomsen, tan16, JESTIE22}. 

The determination of the inductance based on the geometric characteristics, namely the outer and inner side lengths, the width of the conductor, the number of turns and the spacing between them, has been explored in literature \cite{kazimierczuk, wheeler, rosa2, ssmohan}, but for regular polygon shapes (squares, hexagon, etc.) used in RF applications and corresponding to the respective power levels and operating frequencies. Analytical modeling approaches have been reported \cite{aebischer2020, tan16, greenhouse}, which usually conclude to complex mathematical forms that are hard to utilize. In the high-power area (with dimensions up to a few dm) accurate methods have been reported as well \cite{WPT}, estimating the inductance based on magnetic field models of PWs, but resulting in compounded forms.

However, it is still interesting to obtain simple formulas, similar to \cite{wheeler, rosa2, ssmohan} which can estimate the inductance of an RPW through a straight-forward calculation, without compromising the accuracy of the results. This article investigates the optimal modification that adapts three well-known equations, initially developed for low-power square-shaped PWs to RPWs of relevant dimensions for high-power applications. A first approach of this method, discussed in \cite{papadopoulos}, indicated that substituting either the arithmetic or the geometric mean of the two external dimensions in the original equations, provided promising results. This hypothesis was then tested on a few tens of simulated PW samples. In this article, this observation is more thoroughly studied: (i) the optimal combination of the external dimensions is located through the Generalized Mean Value (Power Mean), considering a large variety of possible combinations (p-norm values), and (ii) a greater population of approximately 2600 simulated winding samples is employed, extending the applicability of the results in a wider range of geometric dimensions. The enhancements achieved through this investigation, compared to the original equations are collected in Table \ref{Table:Intro}.

This article is organized as follows: In Section \ref{S_FAI}, an analysis of the original equations is given, discussing the role and the weight of each parameter. In Section \ref{S_PARAM}, the scope of this study is defined, in terms of physical dimensions and power capability, along with the simulation setup used to benchmark the results. The optimization algorithm and the assessment criteria are discussed in Section \ref{S_METH}. The results, confirmed by simulation and laboratory measurements are presented in Section \ref{S_RES}. The Generalized Mean Value or Power Mean (PM) used in the optimization process is further discussed in the Appendix for the sake of completeness.

\begin{table}[]
	\centering
	\caption{Advantages and Limitations of the Three Original and Modified (Proposed) Equations}
	\label{Table:Intro}
\begin{tabular}{lcc}
	\toprule
	& \multicolumn{2}{c}{Equations}                                                                                                                                                            \\ [0.5ex]  
	& Original \cite{ssmohan}            & Modified         \\ [0.5ex]  \hline
	\begin{tabular}[c]{@{}l@{}}Equations\\ Complexity\end{tabular} & Simple   & Simple  \\ [3ex]
	\begin{tabular}[c]{@{}l@{}}Estimation\\ Accuracy\end{tabular}  & \begin{tabular}[c]{@{}c@{}}Good\\ (most < 3\% error)\end{tabular}      & \begin{tabular}[c]{@{}c@{}}Good\\ (most < 3\% error)\end{tabular}      \\ [3ex]
	
	Frequency   & \begin{tabular}[c]{@{}c@{}}Ultra-High-Frequency\\ (RF, in order of MHz)\end{tabular} & \begin{tabular}[c]{@{}c@{}}High-Frequency\\ (Power Applications, \\ in order of kHz)\end{tabular} \\ [3.5ex]
	Rating  & Low Power & High Power                                                                                        \\ [2ex]
	\begin{tabular}[c]{@{}l@{}}Shape\\ Applicability\end{tabular}  & Regular Polygons          & \begin{tabular}[c]{@{}c@{}}Regular and \\Non-Regular\\ Polygons\end{tabular}   \\ \bottomrule                    
\end{tabular}
\end{table}

%Dimensions  & \begin{tabular}[c]{@{}c@{}}Relatively Small\\ (order of \textmu m)\end{tabular}    & \begin{tabular}[c]{@{}c@{}}Relatively Large\\ (order of mm)\end{tabular}      \\ [3ex]

\section{Equation Analysis and Parameter Interpretation}\label{S_FAI}

Several equations have been proposed in literature, and are presented collectively in \cite{kazimierczuk}. In this study, the focus will concentrate on three of them: (i) the empirical Wheeler's equation, which was originally presented for single-layer helical coils \cite{wheeler}, and was later modified to also estimate the inductance of other regular polygon layouts \cite{ssmohan}, (ii) Rosa's equation \cite{rosa2}, which was derived by the current-sheet approximation, and (iii) the Monomial equation \cite{ssmohan}, which was derived by multiple linear regression from a dataset of approximately 19k samples. All equations were developed for square-shaped PWs of small dimensions (in the order of magnitude of $\mu$m), but they have presented accurate results for larger PWs as well (up to a few dm). The three inductance approximation equations are

\begin{figure}[!t]
	\centering
	\includegraphics[width=3.0in]{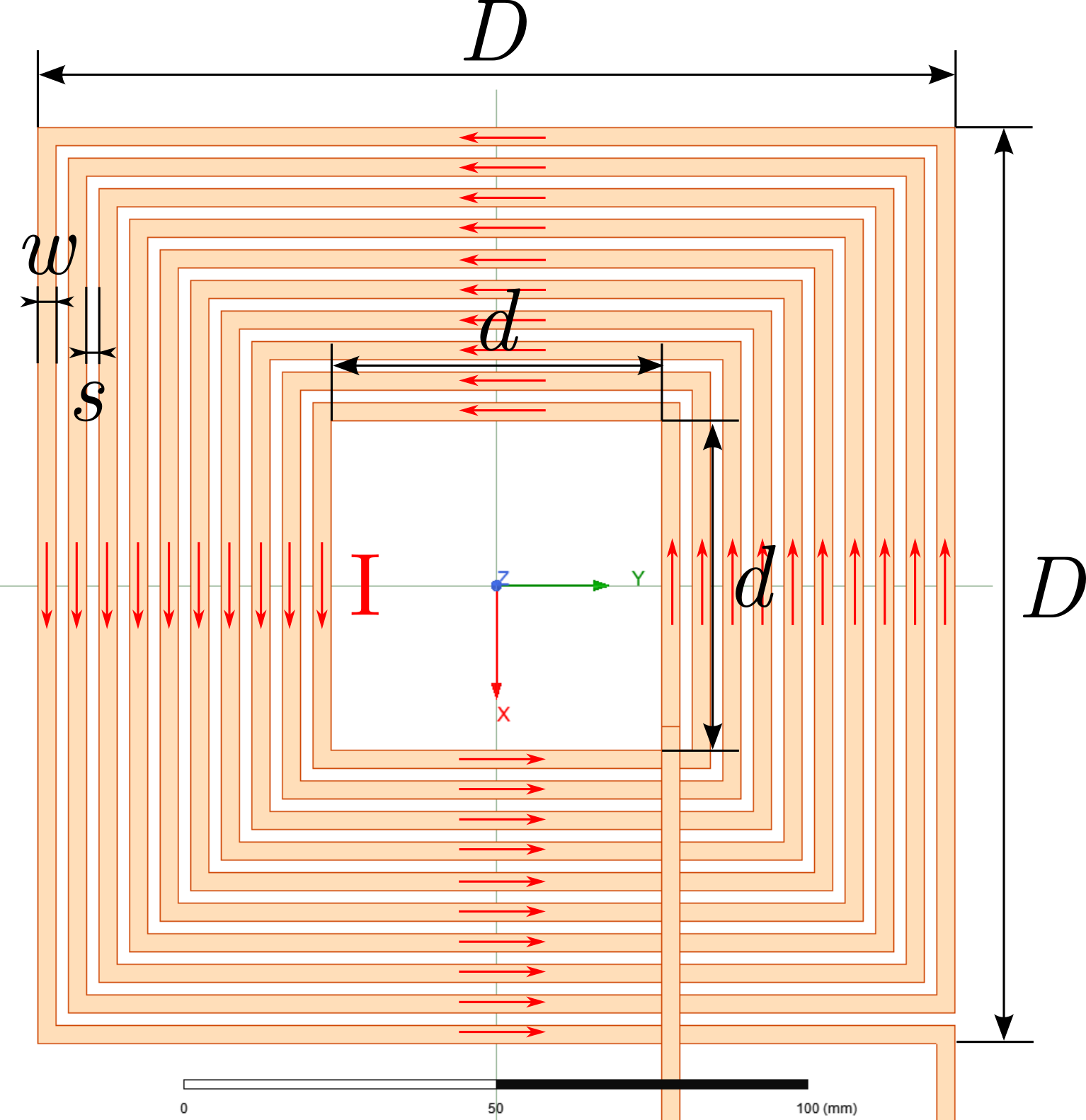}
	\caption{Single-Layer Square Planar Winding of $N= 10$ turns. The external side length $D$, inner side length $d$, copper trace width $w$, and turn spacing $s$ are illustrated. Also, the current direction is presented, which constitutes the basis for the current sheet approximation.}
	\label{fig_parameters}
\end{figure}

{\setlength{\mathindent}{0cm}
\begin{align}
	L_{\text{WH}} = & 1.17 \mu_0 N^2 \frac{D+d}{1+2.75 \rho} \label{eqn_wh}, \\
	L_{\text{RS}} = & \frac{1.27}{4} \mu_0 N^2 (D+d) \underbrace{\left( \ln \left(2.07 \rho \right) + 0.18 \rho + 0.13\rho^2 \right)}_{L(\rho)} \label{eqn_rs}, \\
	L_{\text{MN}} = & 1.54 \mu_0 N^{1.78} \left( \frac{D+d}{2} \right) ^{2.4} D^{-1.21} w^{-0.147} s^{-0.03} \label{eqn_mn},
\end{align}}

\noindent
where $D$ and $d$ are the outer- and inner-side lengths respectively, $\rho = (D-d)/(D+d)$ is the filling factor of the winding, $w$ is the width of the conductive trace, $N$ is the number of turns, and $s$ is the spacing between them. These parameters are presented in Fig. \ref{fig_parameters}, for a square-shaped PW with $N = 10$ turns.

Various observations can be made from (\ref{eqn_wh}) - (\ref{eqn_mn}) regarding the behavior of the inductance. As expected, the inductance increases with the square of the number of turns $N$, except for the Monomial equation, where the coefficient is slightly lower. All equations include a relation to the average side length term $(D+d)/2$: a proportional in the first two cases, and an exponential increase in the third case. This term can be considered a byproduct of the current sheet approximation: traces that run along one direction share a positive mutual inductance with the ones conducting current in the same direction (and are on the same side of the winding). Negative mutual inductance is developed among traces of the opposite sides (conducting current in the opposite direction),  as presented in Fig. \ref{fig_parameters}. Determining the mutual inductance among conductors on the opposite sides of the winding requires the arithmetic mean distance between the two groups, hence, this term appears. Traces that are perpendicular to each other have zero mutual inductance \cite{rosa2}.  

The filling factor $\rho$, which represents the amount of free space in the center of the winding, appears in (\ref{eqn_wh}) and (\ref{eqn_rs}). If $D$ is kept constant, the ratio $\rho$ decreases as $d$ increases. For two windings with the same $D$ and $N$, the one with the greater $d$ is expected to have greater inductance, since its turns are concentrated closer to the outer side, have slightly greater length, and interact more efficiently with their adjacent turns, concluding in higher mutual inductance. Another benefit of windings with small filling factor is the ability to use the central aperture to introduce a ferrite leg, thus reducing the magnetic path resistance of cored components.
According to (\ref{eqn_wh}), where $\rho$ appears in the denominator, increasing $d$ increases the inductance. A similar behavior can be observed in (\ref{eqn_rs}): As $\rho < 1$, the term $L(\rho) = \ln \left(2.07 \rho \right) + 0.18 \rho + 0.13\rho^2$ increases with increasing $d$, as the $(\ln(2.07/\rho))$-term is stronger than $(0.18\rho + 0.13\rho^2)$-term.

In the Monomial approximation, $D$ appears in both the numerator and the denominator. Since the exponent of the first is larger than the second, the total inductance tends to increase with $D$. On the contrary, the total inductance decreases as the width of the trace $w$ and the spacing between the turns $s$ increase, but almost insignificantly. More specifically, the effect of $s$ can be neglected, since it is raised to a power of almost zero. It can be noted that in \cite{ssmohan}, the geometrical parameters of the Monomial are in mm and the resulting inductance in nH. Eq (\ref{eqn_mn}) as it appears here has been modified in order to be in SI values. The first term of 1.54 has been rounded down from 1.542784.

In conclusion, the inductance increases with $N$, $D$, and $d$, which means that for any magnetic component where high inductance is desirable, these terms must receive their greatest possible values. It must be noted that these parameters are not fully-independent: $D$ and $N$ can be considered independent, although restricted by the physical-dimension limitations. The parameters $w$ and $s$ are dictated, however, by the nominal current and voltage of the winding, respectively. The IPC standards 2221, 2222 and 2152 describe the minimum trace width on a rigid PCB as a function of the nominal current and the temperature rise $\Delta T$, as well as the minimum separation between two traces with a specific voltage difference $\Delta V$. The necessary voltage difference between adjacent turns in PWs is discussed in \cite{papadopoulos}. Considering the above, $d$ is dependent on the aforementioned parameters, as in

{\setlength{\mathindent}{2.5cm}
\begin{equation} \label{eqn_d}
	d = D - 2 N (w+s) + 2 s.
\end{equation}}

In typical designs, $w$ is larger than $s$ by one order of magnitude, thus $2 N (w+s) \gg 2 s$, which means that for given values of $N$ and $D$, $d$ is strongly dependent on the sum \mbox{$(w+s)$}. A fast design process can be based on the geometrical parameters of the winding, matching the needs of each specific application.

\section{Parameter Determination} \label{S_PARAM}

\subsection{Windings Physical Dimensions} \label{S_DIMS}

High-frequency power applications present a set of challenges on the design of the magnetic components in general, and more specifically for planar devices. Scaling the design up from RF applications to dimensions that can handle the nominal voltage and current in power applications is required. Furthermore, the introduction of a ferrite core may be necessary to reduce the magnetizing current and contain the magnetic field, which leads to reduced non-active power and electromagnetic interference.

The physical dimensions for high-power planar windings vary from a few mm to a few hundred mm. In this study the outer side dimensions vary from 100 mm to 210 mm, close to the A4 paper size, a PCB size widely available and relatively cheap.

The width of the trace $w$ is determined by the IPC-2221/2 and IPC-2152 standards, using as parameters the maximum current and the acceptable temperature increase $\Delta T$ in the copper. The placement of the traces on external or internal PCB layers must be considered, as the ones enclosed inside the FR-4 can dissipate less heat, hence for a given $\Delta T$ they can carry less current. In this study, the width varies from 3 mm to 5 mm, which for a copper height of \SI{35}{\micro\meter} can handle between 5 A and 7 A, for a $\Delta T $ of $10^o$C, according to SaturnPCB \cite{saturn} which follows the IPC-2152 standard. Note that by relaxing the restriction of $\Delta T = 10^o$C, higher current handling capability is possible. 

Similarly to the above procedure, the spacing between two adjacent turns $s$ is determined using the IPC standards. In this study spacings from 0.1 mm to 2 mm are considered, which correspond to a voltage difference $\Delta V$ from 15 V to 400 V. These values are valid for external uncoated conductors (B2) and can be greatly increased for external coated conductors (B4). It shall be kept in mind that the voltage is not following an equal distribution on each turn, but tends to concentrate more on the first few external turns, as discussed in \cite{papadopoulos}.

The number of turns $N$ vary from 6 to 10 in order achieve two main goals: Firstly, to reach values for the coreless winding inductance in the the order of a few \SI{}{\micro\henry}, which can be further increased to a few mH with the introduction of a core, and secondly, to achieve a relatively large $d$, which not only increases the inductance, but offers a central aperture to accommodate a proper core leg. 

\subsection{Simulation Setup} \label{S_SIM}

\begin{figure}[!t]
	\centering
	\includegraphics[width=3.3in]{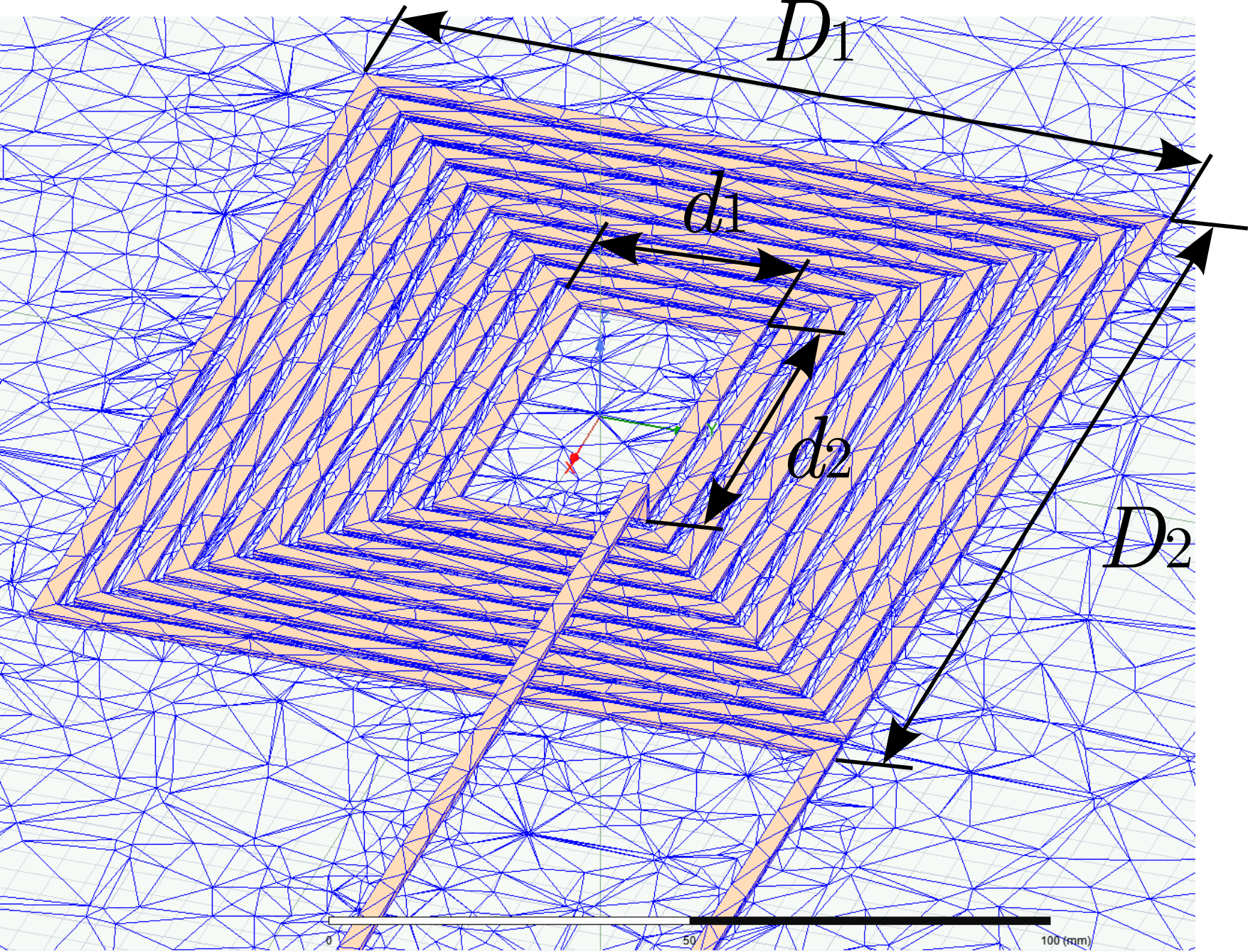}
	\caption{Rectangle Planar Winding of $N = 10$ turns. Outer side lengths $D_1$, $D_2$ and inner side lengths $d_1$, $d_2$ are presented. Tetrahedra meshing in Maxwell3D simulation environment is also presented, for a region of analysis is 700x700x700 mm with around 36k Tets and the coil with around 4k Tets.}
	\label{fig_meshing}
\end{figure}

The proper setup of the Finite Element Method (FEM) simulation is important for the approximation of the inductance and high-frequency effects on the winding. This study considers only the inductance of coreless windings. FEM simulations enable a cheap, automated and relatively quick way to deduce the inductance for a wide variety of parameter values. Rectangle-shaped windings are described by the two outer-side lengths $D_1$ and $D_2$, as presented in Fig. \ref{fig_meshing}. 

The outer-side lengths vary from 100 mm to 210 mm with an increment of 10 mm, $w$ is either $\{3, 4, 5\}$ mm, $s$ is $\{0.1, 0.5, 1, 2\}$ mm, and $N$ is $\{6, 8, 10\}$ turns. The orientation of the winding is irrelevant, hence for a pair $\{D_1, D_2\}$ only one combination is considered. Furthermore, some combinations of $D$, $w$, $s$, and $N$ are impossible to implement as they conclude to a $d<0$ and therefore are neglected. The combinations of the aforementioned parameter values constitute a large dataset, which, excluding the duplicate pairs and the impossible setups, contains 2633 samples.

The region of analysis is defined by a cube with fixed sides of 700x700x700 mm. As it is illustrated in Fig. \ref{fig_meshing}, the generated tetrahedras (Tets) are smaller near and between the copper traces, and become larger as they extend towards the boundaries of the analysis region. On average, 40,000 Tets are generated, 36,000 on the empty space and 4,000 on the coil. The energy error and delta energy were set to 1\%, as an acceptable compromise between accuracy of the model and its runtime. The excitation current was set to 1 A peak sinusoidal wave with a frequency of 100 kHz.

\section{Methodology} \label{S_METH}

\subsection{Optimization Algorithm} \label{S_OPT}

As (\ref{eqn_wh})-(\ref{eqn_mn}) consider only square-shaped PWs, the outer side lengths $D_1$ and $D_2$ of Fig. \ref{fig_meshing}, cannot be directly substituted, but an equivalent mean value is suitable for this use. Based on the Generalized Mean Value or Power Mean (PM), as briefly presented in the Appendix, the parametric mean value of the outer sides is given by $\lVert D \rVert_p = \text{PM}_p(D_1, D_2)$. 

Provided a dataset of $L_{\text{sim}} = f (D, N, w, s)$, an optimization algorithm is required in order to find the optimal parameter $p$ that will provide the best possible results. The function used for this optimization is 

\begin{equation}\label{eqn_error}
	\text{SSE}(p) = \sum_{i=1}^{S} \left( L_{\text{sim},i} - L_{\{\text{WH,RS,MN}\},i}(p) \right) ^2 \text{,} 
\end{equation}

\noindent
where $S$ is the total population of the selected dataset and $i$ is a single sample. The term $\left( L_{sim,i} -  L_{\{\text{WH,RS,MN}\},i}(p) \right) ^2$ is the Squared Error (SE) between the respective equation and the simulation results. The Sum of Squared Errors (SSE) represents the sum of these errors for the entire dataset. For different $p$-values, there is a different combination of $D_1$ and $D_2$ substituted in (\ref{eqn_wh})-(\ref{eqn_mn}), creating different estimation results, indicating different levels of accuracy for each estimation. The parameter $p$ is selected to vary in a range from $[p_{min}, p_{max}]$ with a fixed step increment. The steps of the algorithm are the following:

\begin{enumerate}[i.]
	\item define dataset,
	\item $p = p_{\text{min}}$,
	\item calculate $\lVert {D} \rVert_p$ and $d$ from (\ref{eqn_d}) \label{item_p},
	\item calculate $L_{\text{WH}}$, $L_{\text{RS}}$, and $L_{\text{MN}}$ from (\ref{eqn_wh}) - (\ref{eqn_mn}),
	\item calculate SE $= ( L_{sim,i} - L_{\{\text{WH,RS,MN}\},i}(p) ) ^2$ for each sample $i$, and SSE for the entire dataset from (\ref{eqn_error}),
	\item $p \leftarrow p + \text{increment}$,
	\item if $p$ in range $[p_{\text{min}}, p_{\text{max}}]$ return to (\ref{item_p}), otherwise continue,
	\item find minimum SSE and corresponding $p_{\text{opt}}$
\end{enumerate}

After the optimization process is complete and the optimal $p$ value is obtained, the modified equations of Section \ref{S_RES} can be used directly. As long as the dimensions of the windings are within (or close to) the dataset given in Section \ref{S_DIMS}, the modified equations provide accurate results.

The equivalent inner side length $d$ in step (\ref{item_p}), can either be found from (\ref{eqn_d}), or from $\lVert {d} \rVert_p = \text{PM}_p(d_1,d_2)$. The second way does not offer any significant improvement in the results and will not be considered in the following.

% The intention is not to repeat the optimization process, but use directly the modified equations provided in Section \ref{S_RES}. As long as the dimensions of the windings are within (or close to) the dataset given in Section \ref{S_DIMS}, the modified equations provide accurate results.

\subsection{Assessment Criteria} \label{S_ASSESS}

The optimization algorithm performs an exhaustive search for the optimal $p$ in the predetermined space $[p_{\text{min}}, p_{\text{max}}]$, which is set to $[-5, 5]$ with a step size of $0.001$.  For each $p$, SSE is calculated as in (\ref{eqn_error}) and the $\text{min}|\text{SSE}(p)| = \text{SSE}(p_{\text{opt}})$ is found. In order to illustrate the results, either

\begin{equation}\label{eqn_error2}
	E_p = \sqrt{ \frac{\text{SSE}(p)}{S} }
\end{equation} 

\noindent
or

\begin{equation}\label{eqn_MAE}
	\text{MAE}_p \% = \frac{1}{S}\sum_{i=1}^{S}\frac{|L_{sim,i}-L_{\{\text{WH,RS,MN}\},i}|}{L_{sim,i}} \cdot 100
\end{equation} 

\noindent
can be used. Eq. (\ref{eqn_error2}) essentially presents the RMS error between the simulation results and the equation approximation. While the $\text{SSE}(p)$ value is H$^2$ and does not have any significant physical meaning, $E_p$  presents the root of mean squared error for every sample in a given dataset, measured in \SI{}{\micro\henry}. Equivalently, the results can be illustrated by (\ref{eqn_MAE}), which is dimensionless, but presents the mean absolute error as a percentage, using the simulated inductance result as reference value. While $p_{\text{opt}}$ will be selected, based on the minimization of $E_p$, the results will be presented based on $\text{MAE}_p$, as a normalized value is more effective in visualization.

\section{Simulation and Experimental Results} \label{S_RES}

To easily categorize each dataset, their names represent the parameters that remain the same throughout the optimization process. For example N8w4 represents a dataset, which has a constant number of $N = 8$ turns and traces with $w = 4 $ mm width, while all other parameters vary according to the specifications given in Section \ref{S_SIM}. 

\begin{figure}[!t] 
	\centering
	\subfloat[\label{1a}]{%
		\includegraphics[width=3.5in]{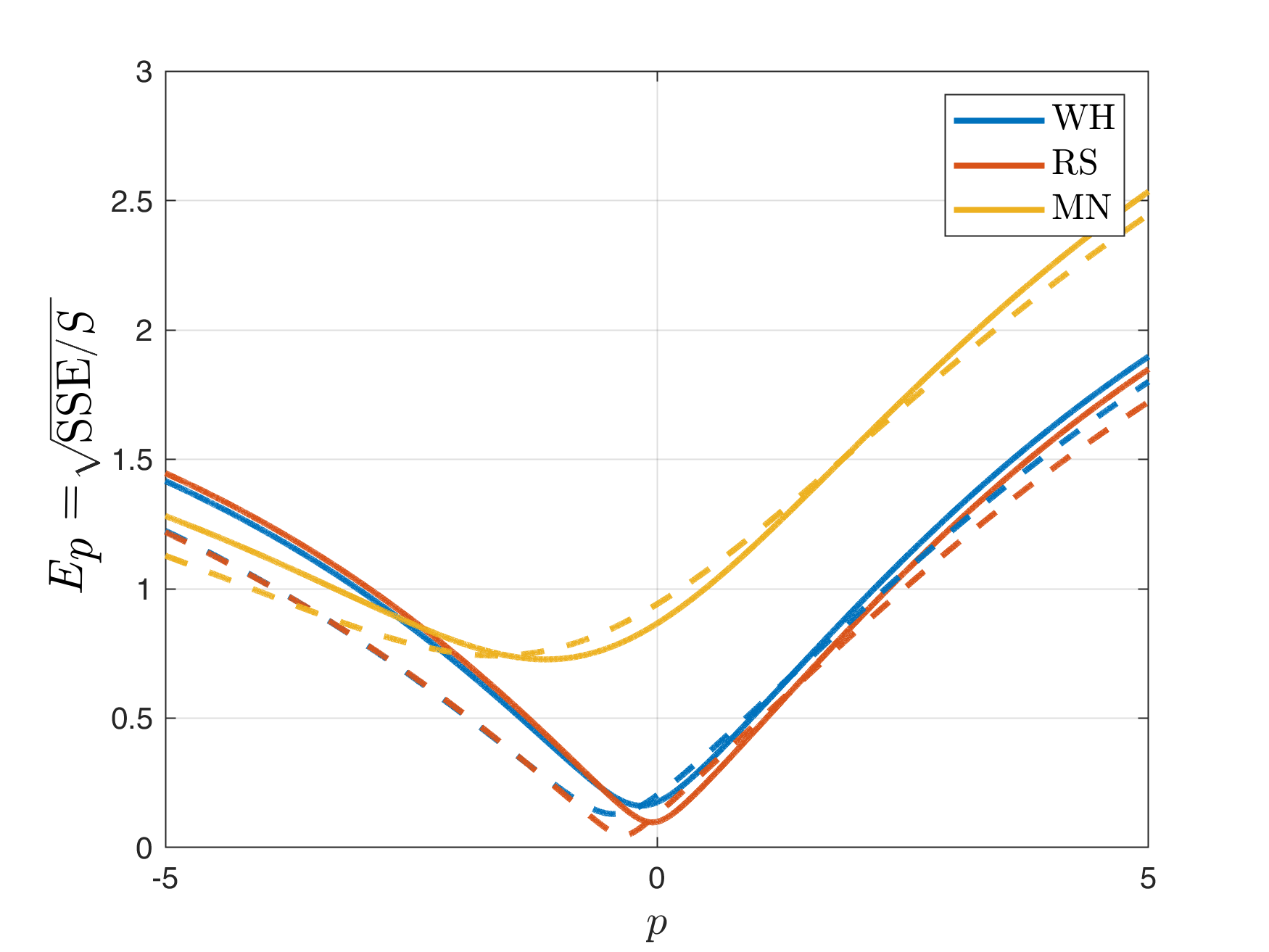}}
	\\
	\subfloat[\label{1c}]{%
		\includegraphics[width=3.5in]{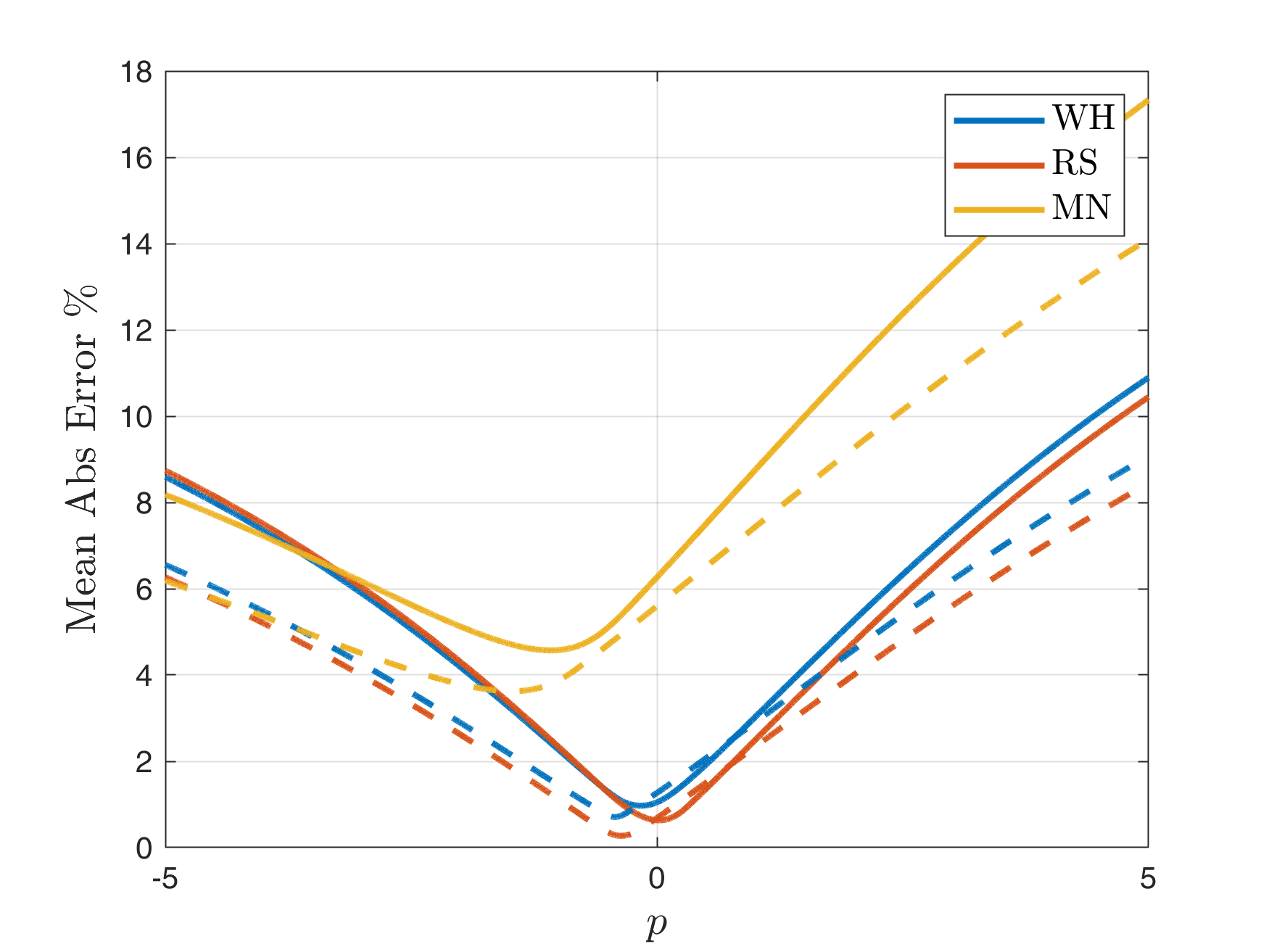}}
	\caption{(a) $E_p$ and (b) MAE\% as a function of $p$ for N10w5 (dashed line) and all samples (continuous line), for each equation. The optimal $p_{\text{opt}}$ depends on the specific dataset but all line present a global minimum, which is in the neighborhood of 0 for Wheeler's and Rosa's, and in the neighborhood of -1 for Monomial.}
	\label{fig_alldata} 
\end{figure}

All datasets present one global minimum (for the search region) with regard to $p$, as illustrated in Fig. \ref{fig_alldata}, where the continuous lines represent the $E_p$ and MAE of all simulated windings, and the dashed lines the respective values for a single dataset (N10w5). The latter is illustrated only as an indicative case; all other datasets behave similarly, with a global minimum around 0 for (\ref{eqn_wh}) and (\ref{eqn_rs}) and around -1 for (\ref{eqn_mn}). 

The Monomial seems to have a slightly different behavior compared to the other two equations, as its MAE is significantly higher for positive $p$ values, and presents the optimal $p$ in a different area. This can be mainly attributed to the coefficient values shown in \eqref{eqn_mn}, which are derived using a linear regression algorithm for a specific dataset of planar inductors, targeting RF applications. It should be noted that recalculating the coefficients of the Monomial equation, based on the dataset of Section \ref{S_SIM}, results in a behavior much more similar to these of Wheeler's and Rosa's, but this is beyond the scope of this article. In any case, only the respective optimal $p$ value should be used for each equation.
	
%, and the behavior of MAE for other $p$ values
%other $p$ values are not meant to be used, since there exist $p$ values that create much better results.}

\addtolength{\tabcolsep}{-1pt} 

\begin{table}[]
	\centering
	\begin{threeparttable}
	\caption{MAE\% and $p_{\text{opt}}$ for different datasets}
	\label{Table:popt}
	\begin{tabular}{rrrrrrr}
		\toprule
		\multicolumn{1}{c}{\multirow{2}{*}{Dataset\tnote{a}}} & \multicolumn{2}{c}{WH}                             & \multicolumn{2}{c}{RS}                             & \multicolumn{2}{c}{MN}                             \\
		\multicolumn{1}{c}{}                          & \multicolumn{1}{c}{$p_{\text{opt}}$} & \multicolumn{1}{c}{MAE\%} & \multicolumn{1}{c}{$p_{\text{opt}}$} & \multicolumn{1}{c}{MAE\%} & \multicolumn{1}{c}{$p_{\text{opt}}$} & \multicolumn{1}{c}{MAE\%} \\ \midrule
		N6w3                 & 0.446                & 1.18                & 0.209                & 0.36                & -1.170               & 4.89            \\
		N6w4                 & 0.160                & 0.54                & 0.214                & 0.34                & -1.198               & 4.95            \\
		N6w5                 & 0.018                & 0.38                & 0.172                & 0.61                & -1.211               & 4.86            \\
		N8w3                 & 0.004                & 0.81                & 0.080                & 0.42                & -1.080               & 4.59            \\
		N8w4                 & -0.170               & 0.74                & -0.011               & 0.43                & -1.104               & 4.32            \\
		N8w5                 & -0.267               & 0.69                & -0.129               & 0.53                & -1.201               & 4.09            \\
		N10w3                & -0.233               & 1.06                & -0.085               & 0.60                & -1.008               & 4.00            \\
		N10w4                & -0.348               & 0.86                & -0.212               & 0.40                & -1.142               & 3.66            \\
		N10w5                & -0.437               & 0.71                & -0.346               & 0.28                & -1.532               & 3.50 			\\
		All Data			 & -0.162				& 0.97				  & -0.049				 & 0.63				   & -1.130     	  	  & 4.58	 		\\	 \bottomrule                  
	\end{tabular}

	\smallskip
	\scriptsize
	\begin{tablenotes}
		\RaggedRight
		\item[a] This column illustrates the parameters that remain constant and their respective values. For example N8w4 represents a dataset which has a constant $N=8$ turns and $w = 4 $ mm, while all other parameters vary. 
	\end{tablenotes}
\end{threeparttable}
\end{table}

\addtolength{\tabcolsep}{1pt}

Table \ref{Table:popt} summarizes ten dataset categories with the corresponding optimal $p_{\text{opt}}$ and the MAE\%, for each equation separately. One observation that can be made for Wheeler's and Rosa's equations is that as $N$ or $w$ increases, $p_{\text{opt}}$ decreases by a small amount, but generally is located in the neighborhood of $0 \pm 0.45 $. For the entire dataset of simulated windings, $p_{\text{opt}}$ is at -0.16 and -0.05, respectively. The MAE is less than 1.2 \% and 0.7 \% for the first and second equations, respectively, values that are within the simulation accuracy. The maximum error observed in the whole dataset for a single simulated winding is \mbox{7.2 \%} for these two equations.

The Monomial, however, does not present a consistent trend regarding the $p_{\text{opt}}$, with respect to $N$ and $w$. Furthermore, its estimation is less accurate compared to the other two equations. Its $p_{\text{opt}}$ is also located elsewhere, i.e., in the neighborhood of $-1.25 \pm 0.25 $, and the MAE is in the range of 3.5\% - 5\%.

\addtolength{\tabcolsep}{-2pt} 

\begin{table}[]
	\centering
	\caption{MAE\% for $p=-1$, $p=0$ and $p=1$ for different datasets}
	\label{Table:p_1}
	\begin{tabular}{rlllllllll}
		\toprule
		\multicolumn{1}{l}{} & \multicolumn{3}{c}{$p = -1$} & \multicolumn{3}{c}{$p = 0$}  & \multicolumn{3}{c}{$p = 1$}                                                \\
		\multicolumn{1}{l}{} & \multicolumn{1}{c}{WH} & \multicolumn{1}{c}{RS} & \multicolumn{1}{c}{MN} & \multicolumn{1}{c}{WH} & \multicolumn{1}{c}{RS} & \multicolumn{1}{c}{MN}   & \multicolumn{1}{c}{WH} & \multicolumn{1}{c}{RS} & \multicolumn{1}{c}{MN}   \\ \midrule
		N6w3        & 3.22    & 2.42    & 5.16    & 1.43    & 0.62    & 6.49    & 1.82    & 1.86    & 8.63    \\
		N6w4        & 2.49    & 2.75    & 5.20    & 0.57    & 0.69    & 6.70    & 2.07    & 1.83    & 9.05    \\
		N6w5        & 2.40    & 3.12    & 5.09    & 0.38    & 0.86    & 6.80    & 2.53    & 2.09    & 9.37    \\
		N8w3        & 2.21    & 2.35    & 4.80    & 0.82    & 0.43    & 6.28    & 2.81    & 2.38    & 8.68    \\
		N8w4        & 2.15    & 2.47    & 4.50    & 0.90    & 0.44    & 6.29    & 3.54    & 2.73    & 9.02    \\
		N8w5        & 2.14    & 2.32    & 4.25    & 0.09    & 0.60    & 6.33    & 3.72    & 2.97    & 9.11    \\
		N10w3       & 2.10    & 2.14    & 4.17    & 1.43    & 0.68    & 5.88    & 4.01    & 3.17    & 8.61    \\
		N10w4       & 1.90    & 1.78    & 3.80    & 1.45    & 0.73    & 5.73    & 3.95    & 3.15    & 8.34    \\
		N10w5       & 1.44    & 1.12    & 3.87    & 1.27    & 0.68    & 5.60    & 2.91    & 2.34    & 7.43    \\
		All Data    & 2.27    & 2.33    & 4.58    & 1.05    & 0.63    & 6.27    & 3.03    & 2.50    & 8.75    \\    
		\bottomrule                  
	\end{tabular}
\end{table}

\addtolength{\tabcolsep}{2pt} 

In order to simplify the calculations for the the inductance estimation, integer $p$ values are preferred, namely $p=-1$ as the optimal $p$ for the Monomial, $p=0$ as the optimal $p$ for Wheeler's and Rosa's equations, and $p=1$ as the most intuitive value (arithmetic mean) which can be used in context of RPWs. The results for these values of $p$ are presented in Table \ref{Table:p_1}. For Wheeler's and Rosa's equations the $\text{MAE}_{p = 0}$ is very close to that of $\text{MAE}_{p = p_{\text{opt}}}$ of each individual case, as it is expected, and in any other case is less than 5 \%. Similarly, the Monomial $\text{MAE}_{p = -1}$ is close to $\text{MAE}_{p = p_{\text{opt}}}$, but $p=0$ still provides relatively accurate results. Eqs (\ref{eqn_wh}) - (\ref{eqn_mn}) can be modified as

%\begin{align}
%	\begin{split}
%		L_{\text{WH}} &= 1.17 \mu_0 N^2 \frac{\lVert D\rVert_0 + \lVert d \rVert_0}{1+2.75 \lVert\rho\rVert_0} 
%		\label{eqn_wh_mod},
%	\end{split}
%\\
%	\begin{split}
%		L_{\text{RS}} = &\frac{1.27}{4} \mu_0 N^2 (\lVert D\rVert_0+\lVert d \rVert_0) \\ 
%		&\left( \ln \left(2.07 \lVert\rho\rVert_0 \right) + 0.18 \lVert\rho\rVert_0 + 0.13\lVert\rho\rVert_0^2 \right) 
%		\label{eqn_rs_mod},
%	\end{split}
%\\
%	\begin{split}
%		L_{\text{MN}} = &1.54 \mu_0 N^{1.78} \left( \frac{\lVert D \rVert_{-1}+\lVert d \rVert_{-1}}{2} \right) ^{2.4} \\ 
%		&\lVert D \rVert_{-1}^{-1.21} w^{-0.147} s^{-0.03} 
%		\label{eqn_mn_mod},
%	\end{split}
%\end{align}

\begin{align}
	L_{\text{WH}} = & 1.17 \mu_0 N^2 \frac{\lVert D\rVert_0 + \lVert d \rVert_0}{1+2.75 \lVert\rho\rVert_0} 
	\label{eqn_wh_mod},	\\
	L_{\text{RS}} = &\frac{1.27}{4} \mu_0 N^2 (\lVert D\rVert_0+\lVert d \rVert_0) \nonumber \\ 
	&\left( \ln \left(2.07 \lVert\rho\rVert_0 \right) + 0.18 \lVert\rho\rVert_0 + 0.13\lVert\rho\rVert_0^2 \right)
	\label{eqn_rs_mod},	\\
	L_{\text{MN}} = &1.54 \mu_0 N^{1.78} \left( \frac{\lVert D \rVert_{-1}+\lVert d \rVert_{-1}}{2} \right) ^{2.4} \nonumber \\ 
	&\lVert D \rVert_{-1}^{-1.21} w^{-0.147} s^{-0.03} 
	\label{eqn_mn_mod},
\end{align}

\noindent
where $\lVert D \rVert_{y}$ is the $y$-norm of the outer sides $D_1$ and $D_2$, $\lVert d \rVert_{y}$ is the $y$-norm of the inner sides $d_1$ and $d_2$, which can be also calculated from (\ref{eqn_d}) since the difference is negligible, and $\lVert\rho\rVert_y = (\lVert D \rVert_{y}-\lVert d \rVert_{y})/(\lVert D \rVert_{y}+\lVert d \rVert_{y})$. The modified equations \eqref{eqn_wh_mod}-\eqref{eqn_mn_mod} can be used directly, for windings within the specified dimensions or in the near vicinity.

\begin{figure}[!htb] 
	\centering
	\subfloat[\label{fig:p_-1}]{%
		\includegraphics[width=3.4in]{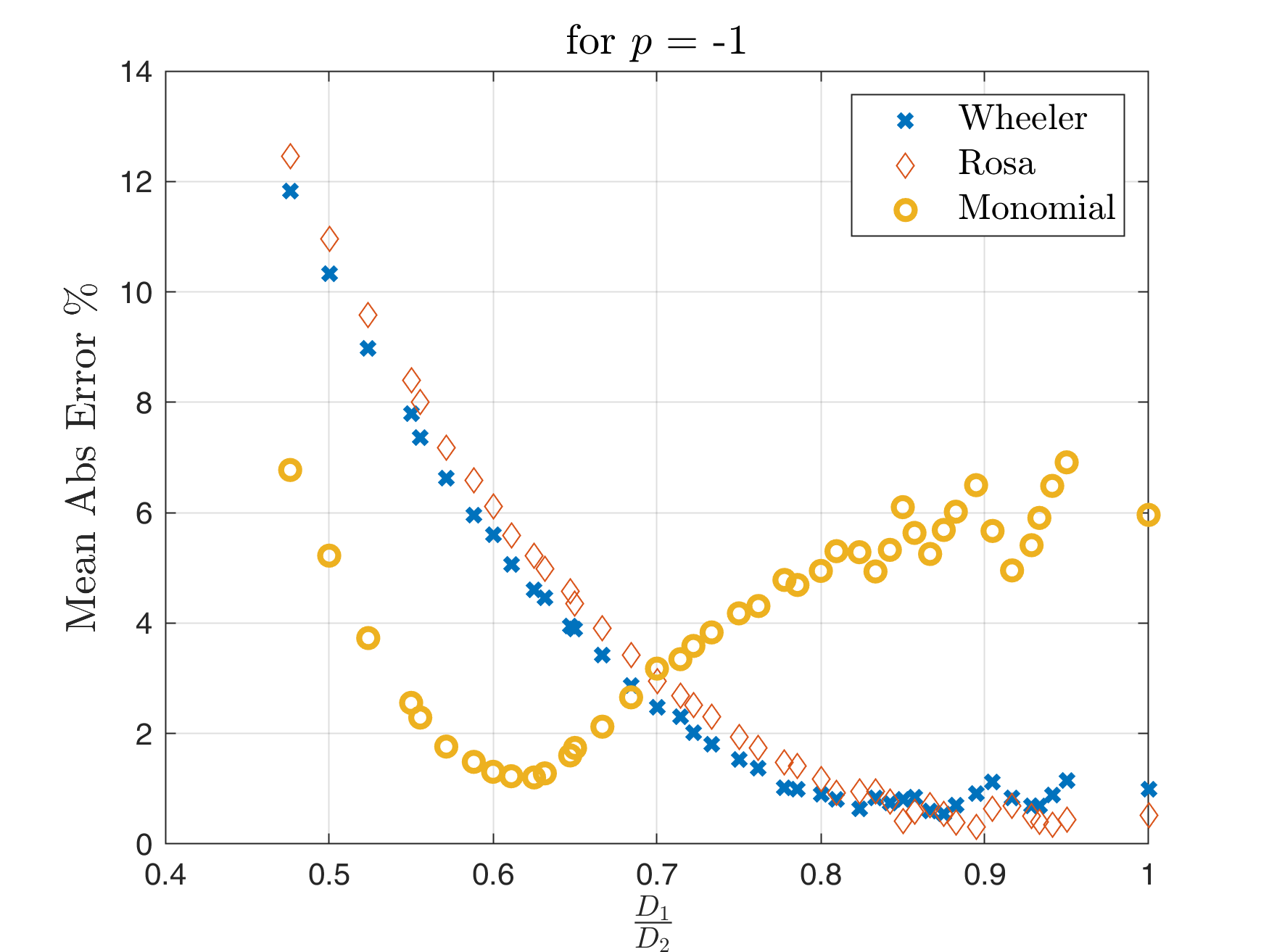}}
	\\
	\subfloat[\label{fig:p_0}]{%
		\includegraphics[width=3.4in]{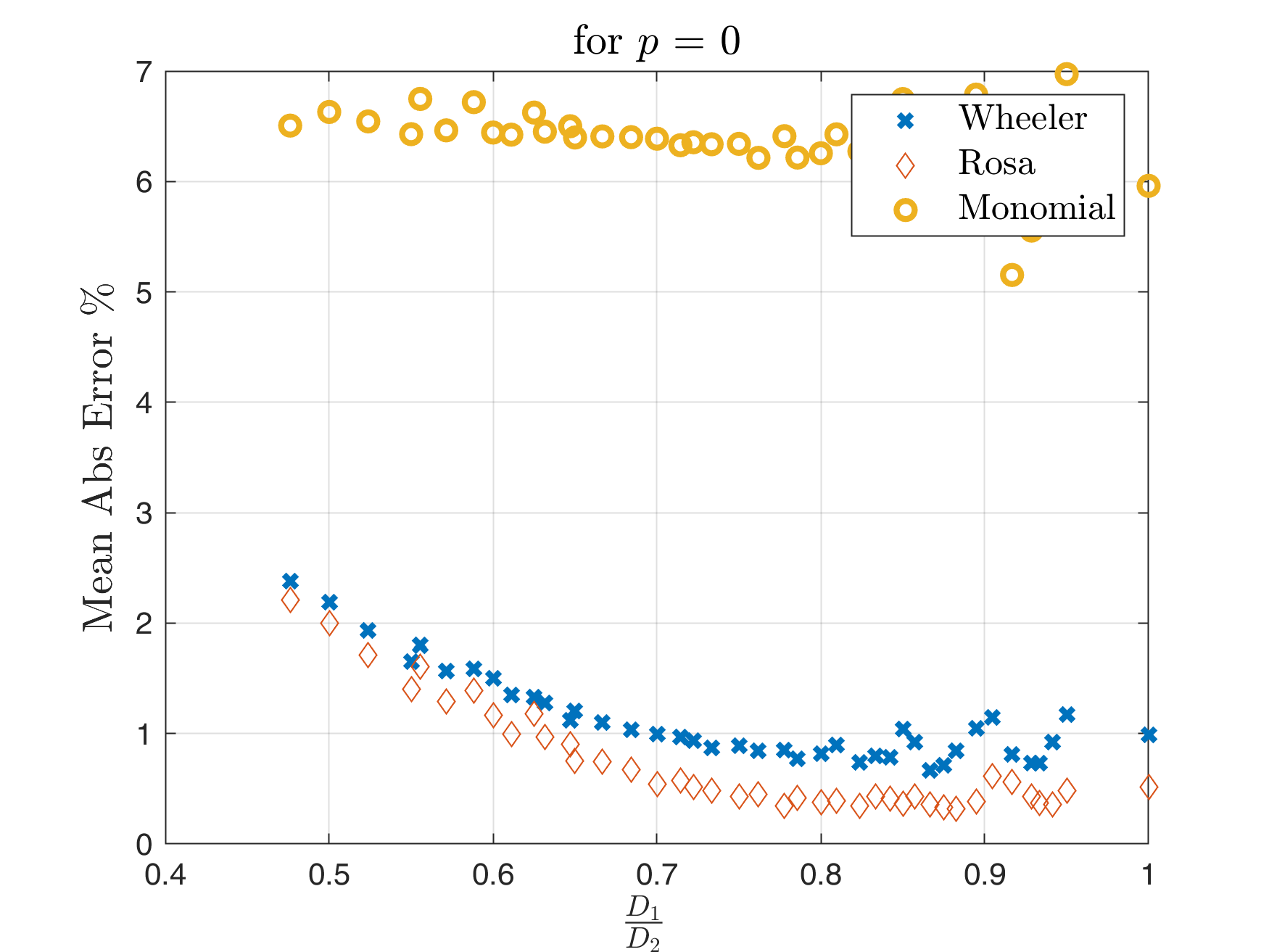}}
	\\
		\subfloat[\label{fig:p_1}]{%
		\includegraphics[width=3.4in]{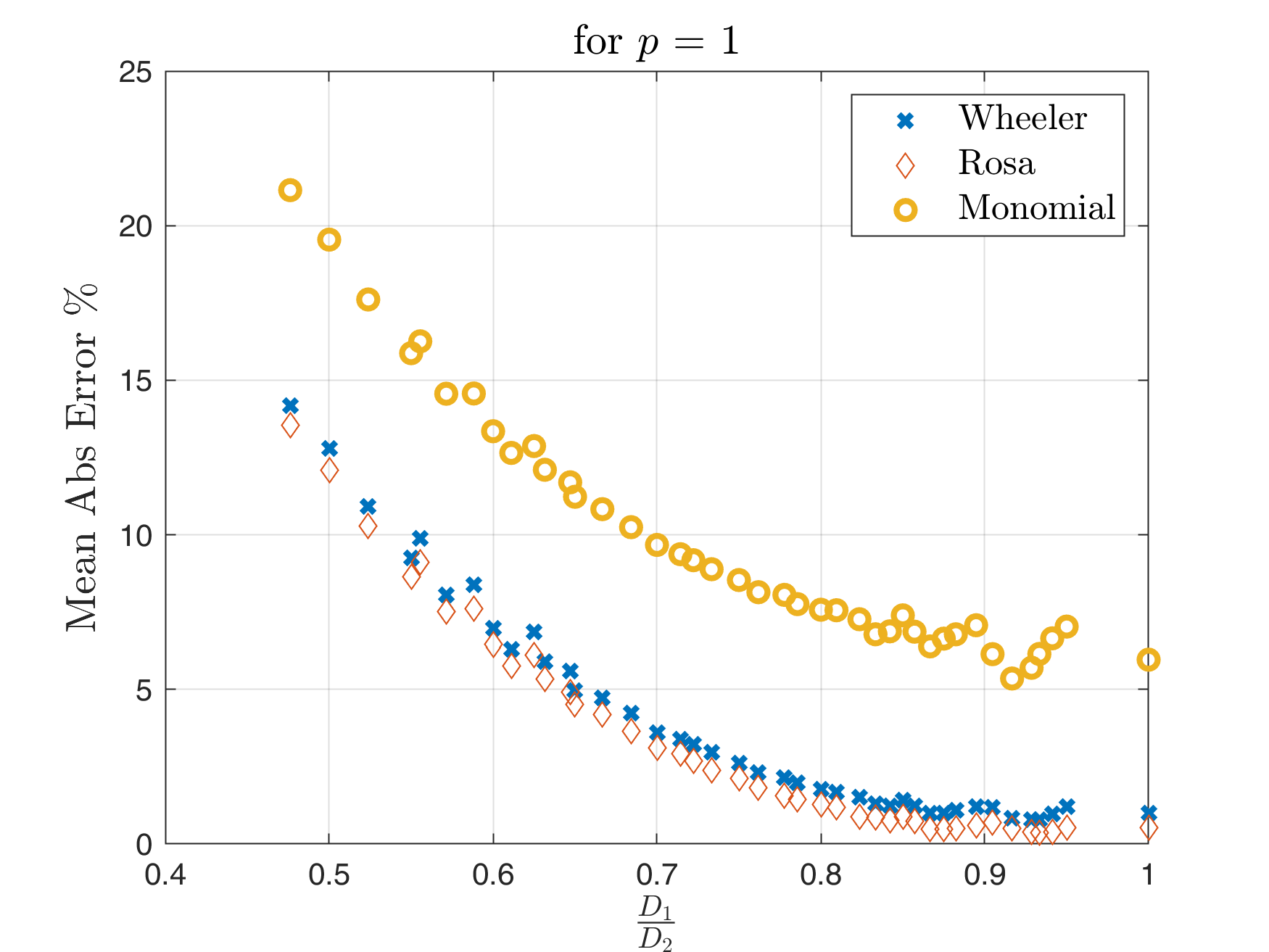}}
	\caption{MAE \% for (a) $p = -1$, (b) $p = 0$, and (c) $p = 1$ as a function of the ratio $D_1 / D_2$, which represents the deformation of the square shape. As $D_1 / D_2$ increases, Wheeler's and Rosa's equation MAE decreases, while the Monomial does not present a consistent behavior.}
	\label{fig:ratio} 
\end{figure}

Having determined the optimal values of $p$ for the adaptation of $D_1$ and $D_2$ values (rectangle outer lengths) to $D$ (square outer length) in each approximation formula, it is important to investigate how accurate each approximation will be for the selected $p$ values, when the outer lengths of the rectangle differ significantly. The deviation from the square shape can be expressed by the ratio $D_1/D_2$, and the MAE\% for the selected $p$ values is shown in Fig. \ref{fig:ratio}, as a function of the $D_1/D_2$ ratio. For Wheeler's and Rosa's equations, the MAE\% for $p = 0$ is significantly lower compared to that for $p = 1$ and $p = -1$, especially when the ratio is small. As it is expected and illustrated in Fig. \ref{fig:ratio}, MAE\% tends to decrease as the ratio increases, i.e., as the windings are getting closer to a square shape. The Monomial approximation does not show the same consistency in its behavior, however, it presents the lowest MAE\% at $p=-1$, as expected from Fig. \ref{fig_alldata}.

\addtolength{\tabcolsep}{-2pt}

\begin{table}[]
	\centering
	\caption{Simulation and Experimental Results}
	\label{Table:SimLab}
	\begin{tabular}{lrrrrrrrrrrrr}
		\toprule
		\multicolumn{1}{c}{\multirow{3}{*}{\parbox[t]{2mm}{\rotatebox{90}{Sample}}}} & \multicolumn{5}{c}{Dimensions}                                                                                                                    & \multicolumn{1}{l}{}         & \multicolumn{1}{l}{}         & \multicolumn{1}{l}{}    \\
		\multicolumn{1}{c}{}                        & \multicolumn{1}{c}{D1}       & \multicolumn{1}{c}{D2}       & \multicolumn{1}{c}{N} & \multicolumn{1}{c}{w}        & \multicolumn{1}{c}{s}        & \multicolumn{1}{c}{Sim}      & \multicolumn{1}{c}{Lab}      & \multicolumn{1}{c}{Er.} \\
		\multicolumn{1}{c}{}                        & \multicolumn{1}{c}{{[}mm{]}} & \multicolumn{1}{c}{{[}mm{]}} & \multicolumn{1}{c}{}  & \multicolumn{1}{c}{{[}mm{]}} & \multicolumn{1}{c}{{[}mm{]}} & \multicolumn{1}{c}{{[}uH{]}} & \multicolumn{1}{c}{{[}uH{]}} & \multicolumn{1}{c}{\%}  \\ \midrule
		\#1          & 100           & 150        & 6     & 4        & 0.1          & 6.159         & 6.174             & 0.24                    \\
		\#2          & 100           & 163        & 8     & 4        & 0.5          & 8.369         & 8.402             & 0.39                    \\
		\#3          & 100           & 163        & 10    & 3        & 0.5          & 13.456        & 13.478            & 0.16                    \\
		\#4          & 210           & 266        & 6     & 5        & 1.0          & 14.553        & 14.396            & -1.08                   \\
		\#5          & 210           & 297        & 10    & 5        & 0.5          & 32.122        & 32.015            & -0.33      \\ \bottomrule                
	\end{tabular}
\end{table}

\addtolength{\tabcolsep}{1pt}

%\addtolength{\tabcolsep}{-3pt} 

\begin{table}[]
	\centering
	\caption{Equations and Experimental Results for the corresponding $p_{\text{opt}}$ of each equation}
	\label{Table:EqLab}
	\begin{tabular}{lrrrrrrr}
		\toprule
		\multicolumn{1}{c}{\multirow{3}{*}{\parbox[t]{2mm}{\rotatebox{90}{Sample}}}} &
		\multicolumn{1}{c}{}         & \multicolumn{2}{c}{Wheeler}                            & \multicolumn{2}{c}{Rosa}                               & \multicolumn{2}{c}{Mono}                               \\
		\multicolumn{1}{l}{} & \multicolumn{1}{c}{Lab}      & \multicolumn{1}{c}{Ind.}     & \multicolumn{1}{c}{Er.} & \multicolumn{1}{c}{Ind.}     & \multicolumn{1}{c}{Er.} & \multicolumn{1}{c}{Ind.}     & \multicolumn{1}{c}{Er.} \\
		\multicolumn{1}{l}{} & \multicolumn{1}{c}{{[}\SI{}{\micro\henry}{]}} & \multicolumn{1}{c}{{[}\SI{}{\micro\henry}{]}} & \multicolumn{1}{c}{\%}  & \multicolumn{1}{c}{{[}\SI{}{\micro\henry}{]}} & \multicolumn{1}{c}{\%}  & \multicolumn{1}{c}{{[}\SI{}{\micro\henry}{]}} & \multicolumn{1}{c}{\%}  \\ \midrule
		\#1 & 6.174                        & 6.145                        & -0.47                   & 6.098                        & -1.25                   & 6.464                        & 4.49                    \\
		\#2 & 8.402                        & 8.424                        & 0.26                    & 8.333                        & -0.83                   & 8.223                        & -2.18                   \\
		\#3 & 13.478                       & 13.575                       & 0.71                    & 13.424                       & -0.40                   & 13.111                       & -2.80                   \\
		\#4 & 14.396                       & 14.421                       & 0.17                    & 14.532                       & 0.94                    & 15.230                       & 5.48                    \\
		\#5 & 32.015                       & 32.479                       & 1.43                    & 32.155                       & 0.44                    & 32.984                       & 2.94   \\ \bottomrule 
	\end{tabular}
\end{table}
%\addtolength{\tabcolsep}{3pt} 

\addtolength{\tabcolsep}{1pt}

Five indicative cases, presented in Table \ref{Table:SimLab}, are selected to be constructed and measured in the lab, in order to evaluate both the simulation results and the equations. The selected dimensions of three windings are within the scope of the simulations, as defined in Section \ref{S_PARAM}, while two of them are outside the dataset used to evaluate the optimal $p$.

\begin{figure}[!t] 
	\centering
	\subfloat[\label{fig:Labsa}]{%
		\includegraphics[width=3.3in]{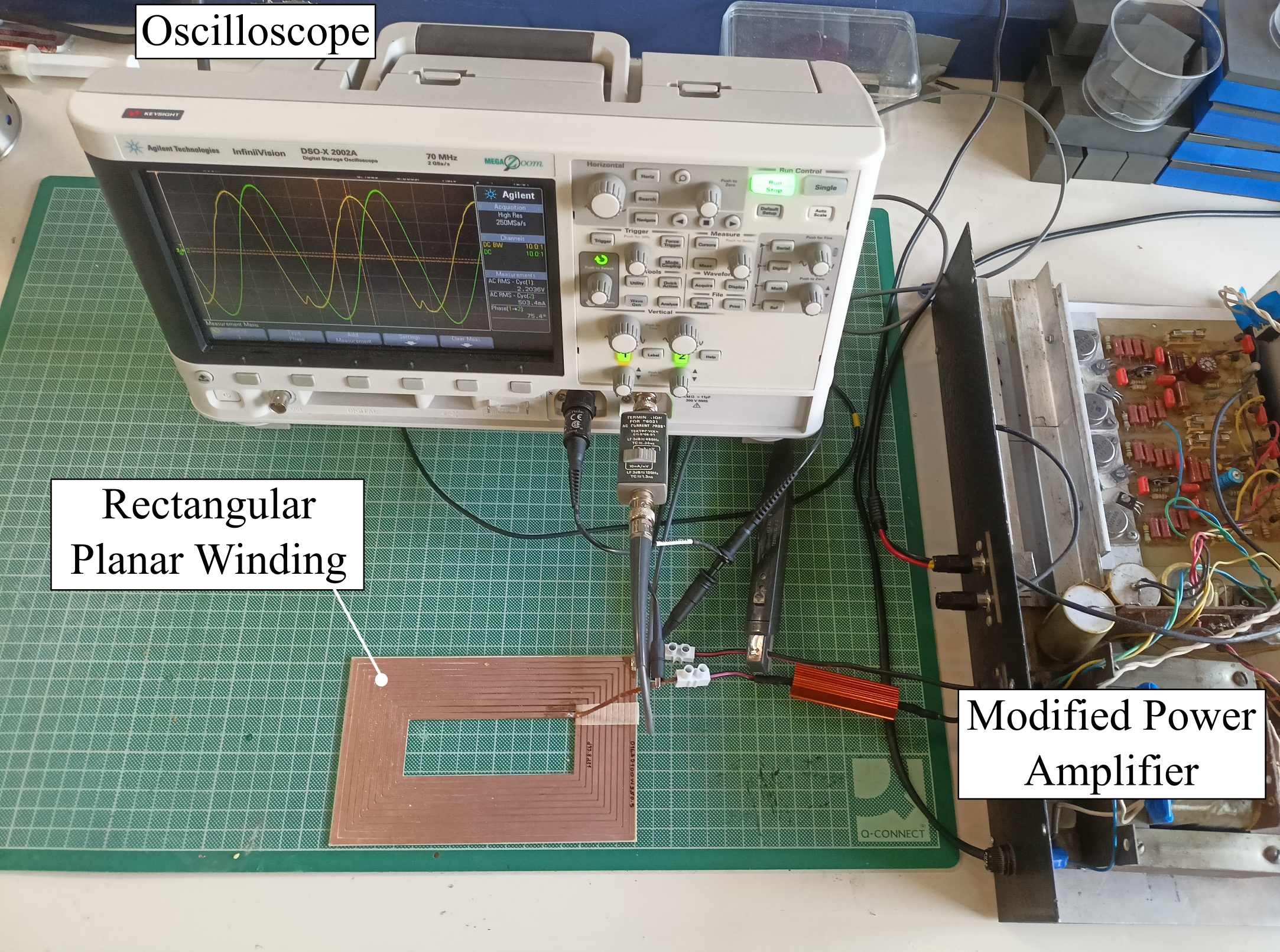}}
	\\
	\subfloat[\label{fig:Labsb}]{%
		\includegraphics[width=3in]{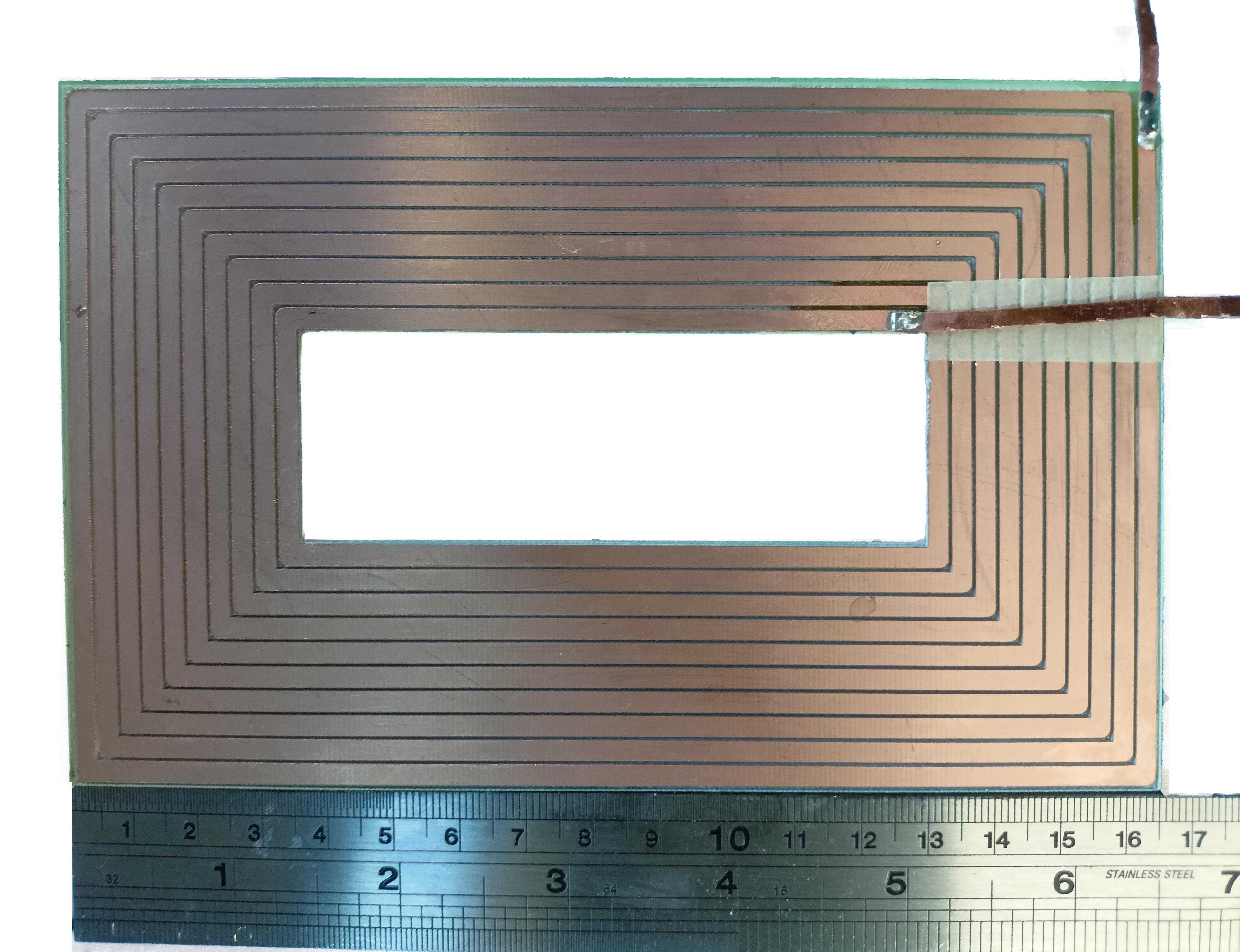}}
	\caption{(a) The experimental setup, presenting the RPW, the modified power amplifier and the oscilloscope via which the inductance is calculated and (b) an indicative printed Single-Layer RPW of 10 turns, with $D_1 = 100$ mm, $D_2 = 163$ mm, $w = 3$ mm, and $s = 0.5$ mm.}
	\label{fig:Labs} 
\end{figure}

A modified power amplifier is used, capable of producing 30 V peak voltage, with frequency up to 50 kHz and a current up to 1.5 A. The laboratory setup is presented in Fig. \ref{fig:Labs} and the inductance is calculated by correlating the voltage excitation to the current response (amplitude and phase difference), for 50 kHz. The error between the simulation and the experimental results is less than 1.1\%, and it is presented in Table \ref{Table:SimLab} in order to evaluate the accuracy of the simulation results. It should be noted that since the self-resonant frequency of the winding is expected to be several orders of magnitude higher than the operating frequency, the inductance will not vary with respect to the latter (up to a reasonable operating frequency, e.g. 100 kHz). 

\begin{figure}[h]
%	\hspace*{-0.5cm}
	\centering
	\includegraphics[width=3.9in]{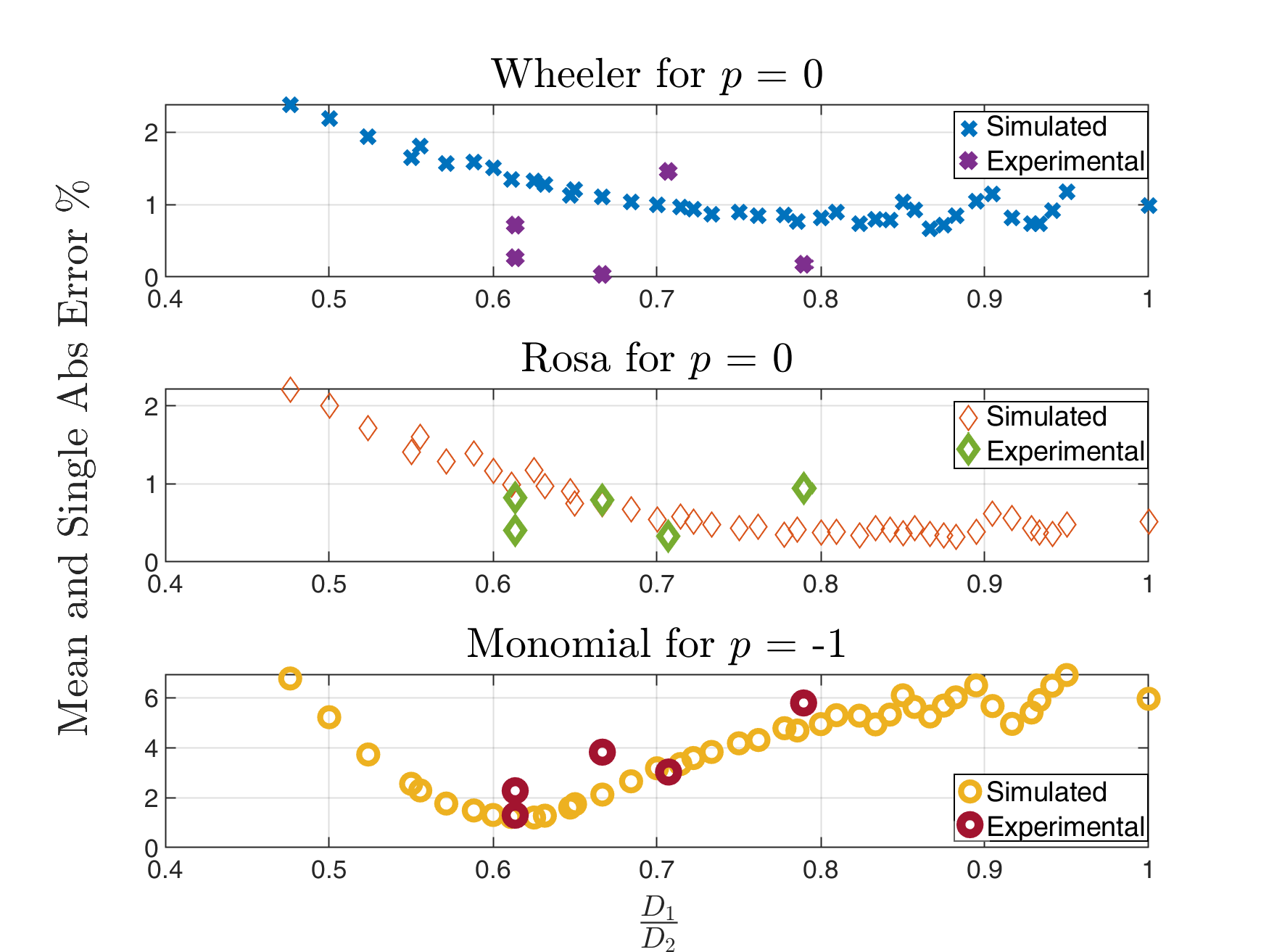}
	\caption{MAE\% for the respective rounded $p_{\text{opt}}$ of each equation, with the corresponding absolute error of the experimental vs. equation results.}
	\label{fig:DRatioExp}
\end{figure}

The comparison between the equations and the experimental results is presented in Table \ref{Table:EqLab}, for the same windings, where the physical dimensions have been omitted. The error is less than 1.5\% for Wheeler's and Rosa's equations and less than 5.5\% for the Monomial, even when considering windings with dimensions outside the initial dataset for which $p$ is optimized. Furthermore, the same comparison is made, based on the ratio $D_1 / D_2$, and as it can be seen in Fig. \ref{fig:DRatioExp}, the absolute error between the measurements and corresponding equation follows the trend of MAE\%. Note that the optimal $p$ value is used for each equation, i.e., $p = 0$ for Wheeler and Rosa, and $p=-1$ for the Monomial, to produce the best possible results. The utilization of other $p$ values is possible, but would increase the estimation error, especially in cases of small $D_1/D_2$, as shown in Fig. \ref{fig:ratio}.

%\begin{figure}[!t]
%\centering
%\includegraphics[width=3.5in]{planar_lab}
%\caption{Indicative case of a Single-Layer Rectangle Planar Winding of 10 turns, with $D_1 = 100$ mm, $D_2 = 163$ mm, $w = 3$ mm, and $s = 0.5$ mm.}
%\label{fig1}
%\end{figure}

\section{Conclusions}

In this study a simple and accurate method for estimating the inductance of high-power rectangle-shaped windings is presented, by extending well-known equations, suitable for regular-polygon shapes. 
This is achieved by utilizing the PM and an optimization method to find the optimal $p$-norm, for which the MAE within a large dataset of simulated windings and equations is minimized. 
For a $p=0$ the MAE for the modified Wheeler's and Rosa's equations is less than 2.5 \%, which corresponds to a small $D_1/D_2$ ratio (a rectangle with two side lengths greatly different). The modified Monomial presents relatively worse results, with its MAE less than 7 \%. The validity of the equations and the simulations is confirmed by the experimental results, on five indicative cases of windings with dimensions even outside the set utilized to obtain the optimal $p$ values. Due to the fact that the Monomial presents relatively worse results, compared to the other two equations, utilizing it with its current coefficients for relatively large windings is not recommended. A recalculation of its power coefficients should be considered to properly adjust the equation to high-power RPWs.

\bibliographystyle{ieeetr}
\bibliography{refs.bib}

\begin{thebibliography}{10}

\bibitem{andersen}
Z.~Ouyang and M.~A.~E. Andersen, ``{Overview of Planar Magnetic
  Technology—Fundamental Properties},'' {\em IEEE Transactions on Power
  Electronics}, vol.~29, no.~9, pp.~4888--4900, 2014.

\bibitem{andersen2}
Z.~Ouyang, O.~C. Thomsen, and M.~A.~E. Andersen, ``{Optimal Design and Tradeoff
  Analysis of Planar Transformer in High-Power DC–DC Converters},'' {\em IEEE
  Transactions on Industrial Electronics}, vol.~59, no.~7, pp.~2800--2810,
  2012.

\bibitem{tria16}
L.~A.~R. Tria, D.~Zhang, and J.~E. Fletcher, ``{Implementation of a Nonlinear
  Planar Magnetics Model},'' {\em IEEE Transactions on Power Electronics},
  vol.~31, no.~9, pp.~6534--6542, 2016.

\bibitem{fletcher}
L.~A.~R. Tria, D.~Zhang, and J.~E. Fletcher, ``{High-Frequency Planar
  Transformer Parameter Estimation},'' {\em IEEE Transactions on Magnetics},
  vol.~51, no.~11, pp.~1--4, 2015.

\bibitem{thermal1}
R.~Shafaei, M.~A. Saket, and M.~Ordonez, ``{Thermal Comparison of Planar Versus
  Conventional Transformers Used in LLC Resonant Converters},'' in {\em 2018
  IEEE Energy Conversion Congress and Exposition (ECCE)}, pp.~5081--5086, 2018.

\bibitem{Litz21}
R.~Yu, T.~Chen, P.~Liu, and A.~Q. Huang, ``{A 3-D Winding Structure for Planar
  Transformers and Its Applications to LLC Resonant Converters},'' {\em IEEE
  Journal of Emerging and Selected Topics in Power Electronics}, vol.~9, no.~5,
  pp.~6232--6247, 2021.

\bibitem{liu21}
S.~Liu, J.~Su, J.~Lai, J.~Zhang, and H.~Xu, ``{Precise Modeling of Mutual
  Inductance for Planar Spiral Coils in Wireless Power Transfer and Its
  Application},'' {\em IEEE Transactions on Power Electronics}, vol.~36, no.~9,
  pp.~9876--9885, 2021.

\bibitem{IPT}
Y.~Wang, H.~Liu, F.~Wu, P.~Wheeler, Q.~Zhou, and S.~Zhao, ``{Research on A
  Three-Coil Hybrid IPT Charger with Improved Tolerance to Coupling Variation
  and Load-Independent Output},'' {\em IEEE Journal of Emerging and Selected
  Topics in Industrial Electronics}, pp.~1--12, 2022.

\bibitem{emin}
E.~Yıldırız, ``{Optimal Design with Generalized Inductance Calculation for
  IPTs Using a Spiral Rectangular Coil Pair},'' {\em Electric Power Components
  and Systems}, vol.~50, no.~19-20, pp.~1212--1222, 2022.

\bibitem{wpt1}
W.~Zhang and C.~C. Mi, ``{Compensation Topologies of High-Power Wireless Power
  Transfer Systems},'' {\em IEEE Transactions on Vehicular Technology},
  vol.~65, no.~6, pp.~4768--4778, 2016.

\bibitem{ev1}
Y.~Park, S.~Chakraborty, and A.~Khaligh, ``{DAB Converter for EV On-Board
  Chargers Using Bare-die SiC MOSFETs and Leakage-Integrated Planar
  Transformer},'' {\em IEEE Transactions on Transportation Electrification},
  pp.~1--1, 2021.

\bibitem{spro}
O.~C. Spro, P.~Lefranc, S.~Park, J.~M. Rivas-Davila, D.~Peftitsis, O.-M.
  Midtgård, and T.~Undeland, ``{Optimized Design of Multi-MHz Frequency
  Isolated Auxiliary Power Supply for Gate Drivers in Medium-Voltage
  Converters},'' {\em IEEE Transactions on Power Electronics}, vol.~35, no.~9,
  pp.~9494--9509, 2020.

\bibitem{margueron1}
X.~Margueron, A.~Besri, P.-O. Jeannin, J.-P. Keradec, and G.~Parent,
  ``{Complete Analytical Calculation of Static Leakage Parameters: A Step
  Toward HF Transformer Optimization},'' {\em IEEE Transactions on Industry
  Applications}, vol.~46, no.~3, pp.~1055--1063, 2010.

\bibitem{buttay}
P.~Demumieux, O.~Avino-Salvado, C.~Buttay, C.~Martin, F.~Sixdenier, C.~Joubert,
  J.~S. Ngoua Teu~Magambo, and T.~Löher, ``{Design of a Low-Capacitance Planar
  Transformer for a 4 kW/500 kHz DAB Converter},'' in {\em 2019 IEEE Applied
  Power Electronics Conference and Exposition (APEC)}, pp.~2659--2666, 2019.

\bibitem{thomsen}
Z.~Ouyang, O.~C. Thomsen, and M.~A.~E. Andersen, ``{Optimal Design and Tradeoff
  Analysis of Planar Transformer in High-Power DC–DC Converters},'' {\em IEEE
  Transactions on Industrial Electronics}, vol.~59, no.~7, pp.~2800--2810,
  2012.

\bibitem{tan16}
W.~Tan, X.~Margueron, L.~Taylor, and N.~Idir, ``{Leakage Inductance Analytical
  Calculation for Planar Components With Leakage Layers},'' {\em IEEE
  Transactions on Power Electronics}, vol.~31, no.~6, pp.~4462--4473, 2016.

\bibitem{JESTIE22}
T.~Yao and W.~Wang, ``{Analysis and Design of Three-End Planar Coupled Inductor
  based DC-DC Converter},'' {\em IEEE Journal of Emerging and Selected Topics
  in Industrial Electronics}, pp.~1--1, 2022.

\bibitem{kazimierczuk}
M.~K. Kazimierczuk, {\em {High-Frequency Magnetic Components}}.
\newblock Wiley, 2~ed., 2013.

\bibitem{wheeler}
H.~Wheeler, ``{Simple Inductance Formulas for Radio Coils},'' {\em Proceedings
  of the Institute of Radio Engineers}, vol.~16, no.~10, pp.~1398--1400, 1928.

\bibitem{rosa2}
E.~B. Rosa, {\em {Calculation of the self-inductances of single-layered
  coils}}, vol.~2.
\newblock 1906.

\bibitem{ssmohan}
S.~Mohan, M.~del Mar~Hershenson, S.~Boyd, and T.~Lee, ``{Simple accurate
  expressions for planar spiral inductances},'' {\em IEEE Journal of
  Solid-State Circuits}, vol.~34, no.~10, pp.~1419--1424, 1999.

\bibitem{aebischer2020}
H.~A. Aebischer, ``{Inductance Formula for Rectangular Planar Spiral Inductors
  with Rectangular Conductor Cross Section},'' {\em Advanced Electromagnetics},
  vol.~9, no.~1, p.~1–18, 2020.

\bibitem{greenhouse}
H.~Greenhouse, ``{Design of Planar Rectangular Microelectronic Inductors},''
  {\em IEEE Transactions on Parts, Hybrids, and Packaging}, vol.~10, no.~2,
  pp.~101--109, 1974.

\bibitem{WPT}
P.~Venugopal, S.~Bandyopadhyay, P.~Bauer, and J.~A. Ferreira, ``{A Generic
  Matrix Method to Model the Magnetics of Multi-Coil Air-Cored Inductive Power
  Transfer Systems},'' {\em Energies}, vol.~10, no.~6, 2017.

\bibitem{papadopoulos}
T.~Papadopoulos and A.~Antonopoulos, ``{Formula Evaluation and Voltage
  Distribution of Planar Transformers Using Rectangular Windings},'' in {\em
  2021 23rd European Conference on Power Electronics and Applications (EPE'21
  ECCE Europe)}, pp.~1--10, 2021.

\bibitem{saturn}
``{Saturn PCB Design Toolkit Version 8.21}.''
  \url{https://saturnpcb.com/saturn-pcb-toolkit/}.
\newblock Accessed: 2022-10-06.

\end{thebibliography}

%Biography text here without a photo.
%\begin{IEEEbiography}[{\includegraphics[width=1in,height=1.25in,clip,keepaspectratio]{fig1.png}}]{IEEE Publications Technology Team}
%In this paragraph you can place your educational, professional background and research and other interests.\end{IEEEbiography}

%\newpage

\clearpage

\appendix[Generalized Mean Value (Power Mean) Optimization for Rectangle-Shaped Windings]\label{S_GMV}

The Generalized Mean Value or Power Mean (PM) is given by

\begin{equation}\label{eqn_GMVfull}
	\text{PM}_p(\vec{x}) = \lVert \vec{x} \rVert_{p} = \left( \frac {1}{N}\sum _{i=1}^{N} x_{i}^p \right)^{\frac{1}{p}} ,
\end{equation}

\noindent where $\vec{x} = [x_1, x_2, \dots, x_N]$ is a vector of dimensions $N \times 1$, and $p$ the $p\text{-norm}$ of $\vec{x}$. For the case of $p = 0$, the PM becomes the well-known geometric mean value, as in

\begin{equation}\label{eqn_GMVGeo}
	\text{PM}_0(\vec{x}) = \lVert \vec{x} \rVert_{0} = \left( \prod_{i=1}^{N} x_{i} \right)^{\frac{1}{N}}.
\end{equation}

\noindent Other well-known and widely used $p\text{-norm}$ examples include: 

\begin{itemize}
	\item[-] $p \rightarrow -\infty$: the minimum value,
	\item[-] $p = -1$: the harmonic mean value,
	\item[-] $p = 1$: the arithmetic mean value,
	\item[-] $p = 2$: the root-mean-square value,
	\item[-] $p \rightarrow \infty$: the maximum value.
\end{itemize}

\noindent For a two-component vector $\vec{x}_{1\times2}$, as is the one composed by two independent outer side lengths $D_1$ and $D_2$, (\ref{eqn_GMVfull}) becomes

\begin{equation}\label{eqn_GMVD1D2}
	\text{PM}_p(D_1, D_2) = \lVert D \rVert_p = \left( \frac {1}{2} \left( D_1^p + D_2^p \right) \right)^{\frac{1}{p}},
\end{equation}

\noindent and (\ref{eqn_GMVGeo})

\begin{equation}\label{eqn_GMVGeoD1D2}
	\text{PM}_0(D_1, D_2) = \lVert D \rVert_0 = \sqrt{D_1D_2}.
\end{equation}

\end{document}